\documentclass[aps,prc,reprint,superscriptaddress,nofootinbib]{revtex4-1}
\usepackage{graphicx}
\usepackage{multirow}
\usepackage{color}
\usepackage{xcolor}
\usepackage{amssymb}
\usepackage{amsmath}
\usepackage{color}
\usepackage{lineno}
\usepackage{enumitem}
\usepackage{comment}
\usepackage[normalem]{ulem}

\newcommand{\br}{{\mathbf r}}

\def\amm#1{\textcolor{purple}{#1}} % AMM comments
\def\nuc#1#2{\relax\ifmmode{}^{#1}{\protect\text{#2}}\else${}^{#1}$#2\fi}

\newcommand{\be}{\begin{eqnarray}}
\newcommand{\ee}{\end{eqnarray}}

\newcommand{\bwt}{\begin{widetext}}
\newcommand{\ewt}{\end{widetext}}

\begin{document}
% \begin{CJK*}{GB}{}
%\preprint{}

\title{Unraveling the reaction mechanisms leading to partial fusion of weakly bound nuclei}
%Alt: Insights into the reaction mechanisms leading to partial fusion of weakly bound nuclei


\author{Jin Lei}
\email[]{jinl@ohio.edu}
%\thanks{}

%\altaffiliation{Present address: Institute of Nuclear and Particle Physics, and Department of Physics and Astronomy, Ohio University, Athens, Ohio 45701, USA}
%\thanks{}
\affiliation{Institute of Nuclear and Particle Physics, and Department of Physics and Astronomy, Ohio University, Athens, Ohio 45701, USA}
%\affiliation{Departamento de FAMN, Universidad de Sevilla, Apartado 1065, 41080 Sevilla, Spain.}

\author{Antonio M. Moro}
\email[]{moro@us.es}
%\thanks{}

\affiliation{Departamento de FAMN, Universidad de Sevilla,
Apartado 1065, 41080 Sevilla, Spain.}


\begin{abstract}
Collisions between complex nuclei may give rise to their total or partial fusion. The latter case is found experimentally to gain importance when one of the colliding nuclei is weakly bound. It has been commonly assumed that the partial fusion mechanism is a two-step process, whose first step is the dissociation of the weakly bound nucleus, followed by the capture of one of the fragments. To assess this interpretation, we present the first implementation of the three-body model of inclusive breakup proposed in the 1980s by Austern \textit{et al.}\  [Phys.\ Rep.\ 154, 125 (1987)] that accounts for both the direct, one-step, partial fusion and the two-step mechanism proceeding via the projectile continuum states. Contrary to the widely assumed picture, we find that, at least for the investigated cases, the partial fusion is largely dominated by the direct capture from the projectile ground-state. 
\end{abstract}

% 25.70.Mn, Projectile and target fragmentation
% 24.10.Eq 	Coupled-channel and distorted-wave models
% 25.45.-z  2H-induced reactions
% 24.87.+y 	Surrogate reactions

%\keywords{}
\date{\today}%
\maketitle

{\it Introduction}.--
\label{sec:intro}
%-------------------------------------------

The understanding of fusion in collisions of composite nuclei  is a problem of utmost importance in various  fields and applications, such as in  reaction networks taking place in astrophysical scenarios \cite{Rol88}, the production of new elements (e.g.~\cite{Oga07,Oga10}), and energy production \cite{Hup19}, among others. 

The first theoretical explanation of fusion started with the seminal work of Bohr \cite{Boh36}, who described the process as the complete merging of the colliding nuclei, giving rise to a  compound nucleus, which eventually dissociates by particle and gamma-ray emission. This appealing picture was soon found to break down in a number of situations. For example, in the 1930s, Oppenheimer and Phillips \cite{OP35} tried to explain the excess of protons in sub-Coulomb deuteron-induced reactions by invoking a partial absorption mechanism, in which only the neutron was captured by the target, favored by the weak-binding and large spatial extension of the deuteron. The idea of partial fusion was revived by Baur and collaborators in the 1970s to account for the large yields of proton singles in deuteron induced reactions at $E_d=25$~MeV on a number of targets \cite{Pam78,Bau80}. The process was described as a two-step
reaction, and coined {\it breakup-fusion} (BF), in which the first step is the breakup of the projectile into $p+n$, and the second step is the absorption of the neutron by the target nucleus. More refined theories were subsequently developed by  Udagawa and Tamura \cite{Udagawa81} and Ichimura, Austern and Vincent (IAV) \cite{IAV85}.
More recently, the BF mechanism has been invoked to explain the phenomenon of complete fusion suppression observed in the above-barrier nuclear collisions with weakly bound nuclei, such as $^{6,7,8}$Li and $^{9}$Be \cite{Das99,Tri02,Das02,Das04,Muk06,Rat09,LFC15}. This suppression amounts up to $\sim$30\% for these nuclei, is roughly independent of the target nucleus and is typically accompanied by significant yields of evaporation products compatible with the partial absorption of the projectile, also referred to as {\it incomplete fusion}, ICF.   However, some recent experimental results \cite{Coo19} suggest that the ICF products are compatible with a direct, one-step mechanism, thus putting into question the BF picture.

From the theoretical point of view, the situation is also unclear. Different  models have been proposed to account for this CF suppression and the related ICF cross sections, including classical  \cite{Dia07,Coo16}, semiclassical \cite{Mar14,Kol18} and quantum-mechanical \cite{Dia02} approaches. Most of them exploit the two-step, breakup-fusion picture. Although in most calculations the coupling to the breakup channels was found to produce a reduction of CF, the predicted suppression is systematically too small. 

In a recent work \cite{Jin19}, we presented a novel approach which provides CF and ICF cross sections within a common framework. Furthermore, the  model was able to account for the observed CF suppression in the $^{6,7}$Li+$^{209}$Bi reactions, for a wide range of incident energies.  Despite the good agreement with the data, the calculations of \cite{Jin19} were not able to answer the important question on whether the ICF proceeds as a two-step process, as assumed by the BF picture, or it is actually a one-step mechanism. The reason is that those calculations were done with the DWBA version of the IAV model. As such, the entrance channel was described with an effective optical potential reproducing the corresponding elastic scattering data. {Although the success of the DWBA approximation to explain these and other inclusive breakup data suggests the dominance of the one-step mechanism over the BF mechanism, the fact that the entrance channel optical potential used in DWBA is commonly adjusted to reproduce the elastic scattering data implies that this potential may implicitly  include breakup contributions, corresponding to situations in which the projectile dissociates prior to its total or partial absorption by the target, which correspond to the first step of the BF mechanism.} 

It is the goal of this work to elucidate the nature of the ICF process and, in particular, to assess the validity of the BF picture. For that,  one needs a model which incorporates explicitly the intermediate breakup channels of the projectile. Such a model was in fact put forward by Austern {\it al.}~\cite{Aus87} in a three-body version of the IAV theory, in which the entrance channel wavefuction was described  using an expansion in projectile eigenstates. This three-body wavefunction is identical to that used in the continuum-discretized coupled-channels (CDCC) method so we will refer to this extended IAV model as IAV-CDCC.  This IAV-CDCC  has not been applied in practice due to its numerical complexity.

In this work we present the first implementation of the IAV-CDCC theory and apply it to several reactions induced by weakly bound projectiles.  In addition to disentangling the nature of the ICF process, this study will serve to assess the accuracy of the commonly adopted DWBA approximation of the IAV model.

\begin{figure}[tb]
\begin{center}
 {\centering \resizebox*{0.96\columnwidth}{!}{\includegraphics{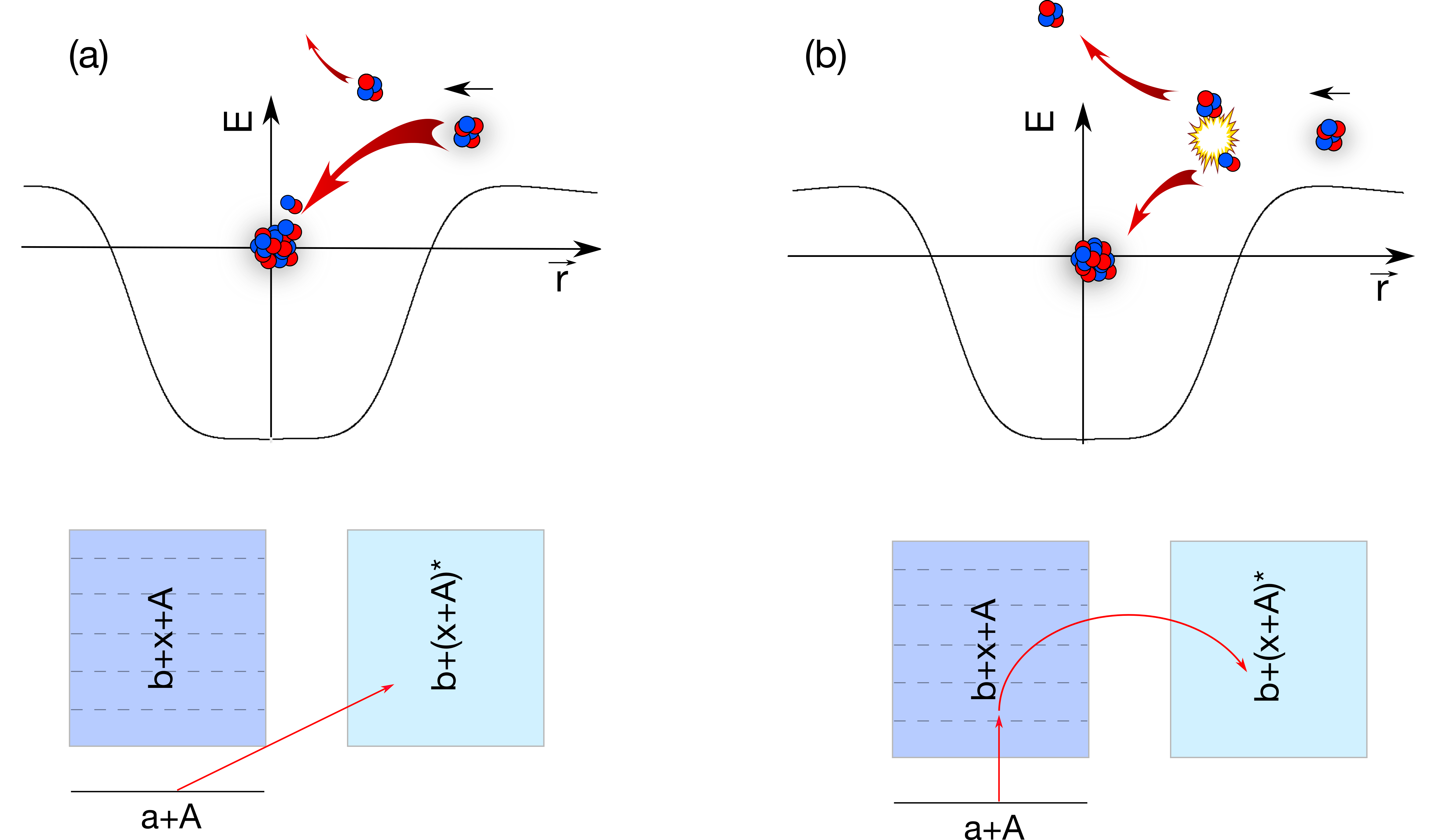}} \par}
\caption{\label{fig:NEB_path}Illustration of the direct (left) and two-step (right) paths leading to partial capture of the projectile. See text for details.}
\end{center}
\end{figure}

%\bigskip
%------------------------------------------------------------------------
{\it  Theoretical framework}.--
 \label{sec:theo}
%--------------------------------------------------------
We consider a process in which a two-body projectile $a=b+x$  collides with a target nucleus $A$, emitting the fragment $b$. Schematically,  
\begin{equation}
\label{eq:1step}
a(=b+x)+A \to  b+ B^*, 
\end{equation}
where $B^*$ denotes any possible final state of the $x+A$ system. This includes the elastic breakup (EBU) process, in which both $b$ and $x$ scatter elastically from $A$, and hence the latter is left in its ground state. The other contributors, which we  call globally non-elastic breakup  (NEB),  are those in which $x$ undergoes a non-elastic interaction with the target, including $x+A$ inelastic scattering, nucleon exchange between $x$ and $A$ and fusion. The latter corresponds to the incomplete fusion (ICF) process mentioned in the introduction. 

The ICF is usually interpreted a two-step process \cite{Udagawa80,Udagawa81,Boselli2016,Tripathi05,Shrivastava13,Utsunomiya75}. For a two-body weakly bound projectile $a$ with a target $A$, such a process may symbolically be written as 
\begin{equation}
\label{eq:2step}
%a(=b+x)+A\to b+x + A \to  b+ B^*,
a+A\to b+x + A \to  b+ B^* .
\end{equation}
In this picture, the projectile is first excited into its continuum states and then one of the fragments ($x$ in this case) {is absorbed by the target}. However, the same final state can in principle be reached via the direct, one-step process in which the $x$ fragment is directly absorbed by the target nucleus, without the intermediate breakup states, as implied by recent experimental results \cite{Dia07,Coo16}. This process is possible invoking for example a {\it Trojan Horse} (TH) mechanism \cite{Jin19}.
These two possible scenarios are depicted in Fig.~\ref{fig:NEB_path}. 

To disentangle the nature of ICF, we make use of the three-body theory proposed by Austern {\it et al.}~\cite{Aus87} (the IAV-CDCC model referred in the introduction), in which the NEB cross section for the inclusive process  $A(a,b X)$ is given by the closed-form formula
\begin{equation}
\label{eq:iav_3b}
\left . \frac{d^2\sigma}{dE_b d\Omega_b} \right |_\mathrm{NEB} = -\frac{2}{\hbar v_{a}} \rho_b(E_b)  \langle \varphi_x (\mathbf{k}_b) | \mathrm{Im}[U_{xA}] | \varphi_x (\mathbf{k}_b) \rangle   ,
\end{equation}
where $\rho_b(E_b)$ is the density of states of the particle $b$,
%=k_b \mu_{b} /[(2\pi)^3\hbar^2]$, 
$v_a$ is the velocity of the incoming particle, $U_{xA}$ is the optical potential describing $x+A$ elastic scattering, and  $\varphi_x(\mathbf{k}_b,\br_{xA})$ is a relative wave function describing the motion between $x$ and $A$ when particle $b$ is scattered with momentum $\mathbf{k}_b$. This function is obtained 
from the equation
\begin{equation}
\label{eq:inh}
\varphi_x(\mathbf{k}_b,\br_{x}) =\int G_x (\br_{x},\br'_{x}) \langle \br'_{x}\chi_b^{(-)}| V_\text{post}|\Psi^{3b(+)} \rangle d\br'_{x}
\end{equation}
 where $G_x$  is the Green's function with optical potential $U_{xA}$,
$\chi_b^{(-)*}(\mathbf{k}_b,\br_{b})$ is the distorted wave describing the relative motion between $b$ and $B^*$ compound system (obtained with some optical potential $U_{bB}$),  $V_\mathrm{post} \equiv V_{bx}+U_{bA}-U_{bB}$ is the post-form transition operator and $\Psi^{3b(+)}$ the three-body scattering wave function. Note that the imaginary part of $U_{xA}$  accounts for all non-elastic processes between $x$ and $A$ and hence Eq.~(\ref{eq:iav_3b}) includes the ICF as well as other NEB contributions. Further details can be found in Ref.~\cite{Jin15}. 

The {\it exact}  wave-function $\Psi^{3b(+)}$  appearing in Eq.~(\ref{eq:inh}) could  in principle be obtained by solving the Faddeev equations~\cite{Faddeev}. However, due to its numerical complexity and to the non-trivial definition of the three-body boundary condition~\cite{Deltuva2014}, Austern {\it et al.}~\cite{Aus87}
proposed as an alternative approximating this three-body wavefunction by an expansion in terms of $b+x$ states, including continuum components, i.e.,
 \begin{align}
 \label{eq:3bwf}
  \Psi^\mathrm{3b(+)} (\mathbf{r}_a,\mathbf{r}_{bx}) & = 
  \sum_{i}\phi_a^{i}(\mathbf{r}_{bx})\chi_a^{i(+)}(\mathbf{r}_a) 
  \nonumber \\ 
 & +\int d \mathbf{k} ~ \phi_a(\mathbf{k},\mathbf{r}_{bx})\chi_a^{(+)}(\mathbf{K},\mathbf{r}_a),
 \end{align}
 where $\{ \phi_a^i(\mathbf{r}_{bx}), \, \phi_a(\mathbf{k},\mathbf{r}_{bx}) \}$ are the eigenfunctions of the projectile Hamiltonian for bound and continuum states, respectively, with $i$  a discrete index for projectile bound  states, and  $k$ the asymptotic momentum of $b+x$ scattering states. The {\it distorted waves}  $\{ \chi_a^{i(+)}(\mathbf{r}_a),\, \chi_a^{(+)}(\mathbf{K},\mathbf{r}_a)\}$ describe the projectile-target relative motion for each projectile state. For continuum states, these functions depend on the momentum $K$, which is related to the internal momentum $k$ by energy conservation. To make (\ref{eq:3bwf}) calculable, the integral over continuum states is 
 approximated by a discrete expansion in a basis of square-integrable functions, as done in the so-called continuum-discretized coupled-channels (CDCC) method \cite{Aus87,Tho09},
\begin{align}
\label{eq:cdccwf}
 \Psi^\mathrm{3b(+)}   & \simeq \Psi^\mathrm{CDCC(+)} (\mathbf{r}_a,\mathbf{r}_{bx})   
= \sum_{i}\phi_a^{i}(\mathbf{r}_{bx})\chi_a^{i(+)}(\mathbf{r}_a)   \nonumber \\
& + \sum_{c}^{N} \phi_a^{c}(k_c,\mathbf{r}_{bx})\chi_a^{c(+)}(K_c,\mathbf{r}_a),
\end{align}
where $c=\{n,j,m\}$, with  $j,m$ the angular momentum and projection of the continuum states and $n$  a discrete index labelling the discretized continuum states. The maximum angular momentum $j$ and wavenumber $k$ is determined by convergence of the studied observables. In the present calculations, we adopt the standard binning method \cite{Aus87,Tho09}, in which the discretized continuum states are represented by wave packets built upon superposition of the $b+x$ scattering states for predefined energy intervals ({\it bins}). The widths of these bins must be chosen small enough so as to produce converged elastic and breakup observables. The radial functions $\chi_a^{i(+)}(\mathbf{r}_a)$ and $\chi_a^c(K_c,\mathbf{r}_a)$ are obtained by solving a system of coupled-differential equations \cite{Aus87,Tho09}.

Inserting the CDCC wave function (\ref{eq:cdccwf}) into Eq.~(\ref{eq:inh}) yields a full three-body description of NEB cross sections. In addition, one can isolate the direct, one-step mechanism contribution by retaining only the ground-state component of Eq.~(\ref{eq:cdccwf}) in Eq.~(\ref{eq:inh}). This approximation will be referred to as IAV-CDCC(gs) in the calculations presented below. 

We conclude this section by noting that one could in principle estimate the ICF content of the NEB cross section by splitting in Eq.~(\ref{eq:iav_3b}) the potential $U_{xA}$ into an inner part and a peripheral one, with the former accounting for the ICF \cite{Mas90,Udagawa85,Ber17}. We prefer however to focus the discussion on the full NEB  to avoid the ambiguity inherent to this separation.

\medskip

%---------------------------------------------------------
{\it Application to the deuteron and $^6$Li induced reactions}.--
%------------------------------------------------------
We first consider the  breakup reaction $^{93}$Nb($d$,$pX$) at $E_d=25.5$ MeV. This reaction  was already analyzed in our previous work~\cite{Jin15} with the DWBA version of the IAV model, finding a good agreement with experimental data.
%. Although the validity of DWBA is not tested at that time, the calculations give a reasonable account of the experimental data. 

Here we compare the NEB differential cross sections using the IAV model, with different choices for the $\Psi^{3b(+)}$  wave-function in  Eq.~(\ref{eq:inh}), namely, the DWBA approximation (IAV-DWBA), the full CDCC wave-function (IAV-CDCC) and the truncated CDCC wave-function, in which only the g.s.\ component   of (\ref{eq:cdccwf}) is retained in Eq.~(\ref{eq:inh}) (IAV-CDCC(gs)). We adopt the same potentials  used in our previous calculations. For the CDCC calculations, the $n-p$ states were included for $\ell=0-4$ partial waves and up to a maximum excitation energy of 20~MeV. For the DWBA results, the deuteron-target potential is taken from Ref.~\cite{Han06} and the potential depth is adjusted to reproduce the elastic scattering differential cross section computed by CDCC. This procedure is intended to reduce uncertainties when comparing NEB differential cross section calculated by these methods. 
To simplify the calculations, we ignore intrinsic spins.

%-------------------------------------------------------------
\begin{figure}[tb]
\begin{center}
 {\centering \resizebox*{0.85\columnwidth}{!}{\includegraphics{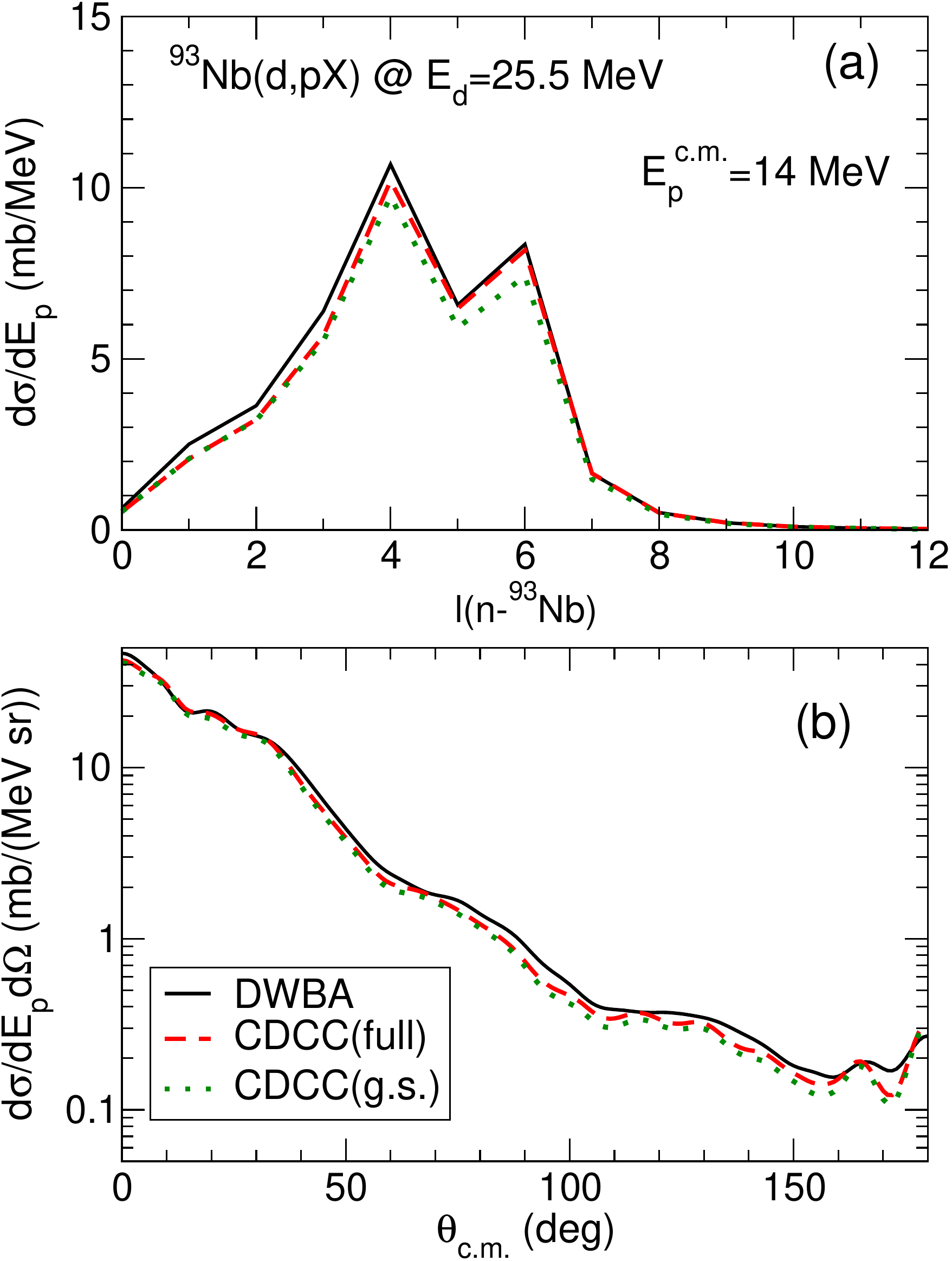}} \par}
\caption{\label{fig:d93nb} Non-elastic breakup contribution for the reaction for $^{93}$Nb($d$,$p$X) at $E_\text{lab}=25.5$~MeV for an outgoing proton C.M.\ energy of $14$~MeV. (a) Energy differential cross section as a function of the neutron-target orbital angular. (b) Double differential cross section angular distribution.}
\end{center}
\end{figure}
%-------------------------------------------------------------

In Fig.~\ref{fig:d93nb}(a), we show the calculated angle-integrated NEB differential cross section, $d\sigma/dE_p$ as a function of the neutron-target orbital angular momentum corresponding to a proton energy of $E_p=14$ MeV in the C.M. frame. 
%The solid, dashed, dotted lines are respectively 
%the results by using DWBA, CDCC, and only ground state of CDCC wave %functions. 
%\sout{The solid line is the DWBA calculation. The dashed line is the result by using the full CDCC wave functions. The dotted line is obtained by using the ground state part of the full CDCC functions.}. 
The solid, dashed, and dotted lines correspond, respectively, to the IAV-DWBA, IAV-CDCC and IAV-CDCC(gs) calculations. We find that all these three calculations give {very similar results.} 
In Fig.~\ref{fig:d93nb}(b) we show the results for the double differential cross section angular distributions ($E_p=14$~MeV).  The three calculations give essentially the same angular shape, with only minor differences seen at the larger angles. {These calculations clearly indicate that, for this reaction, the NEB processes (including ICF) take place directly from the projectile ground state, contrary to the BF picture, and that the BF mechanism is marginal.}
 
%-------------------------------------------------------------
\begin{figure}[tb]
\begin{center}
 {\centering \resizebox*{0.85\columnwidth}{!}{\includegraphics{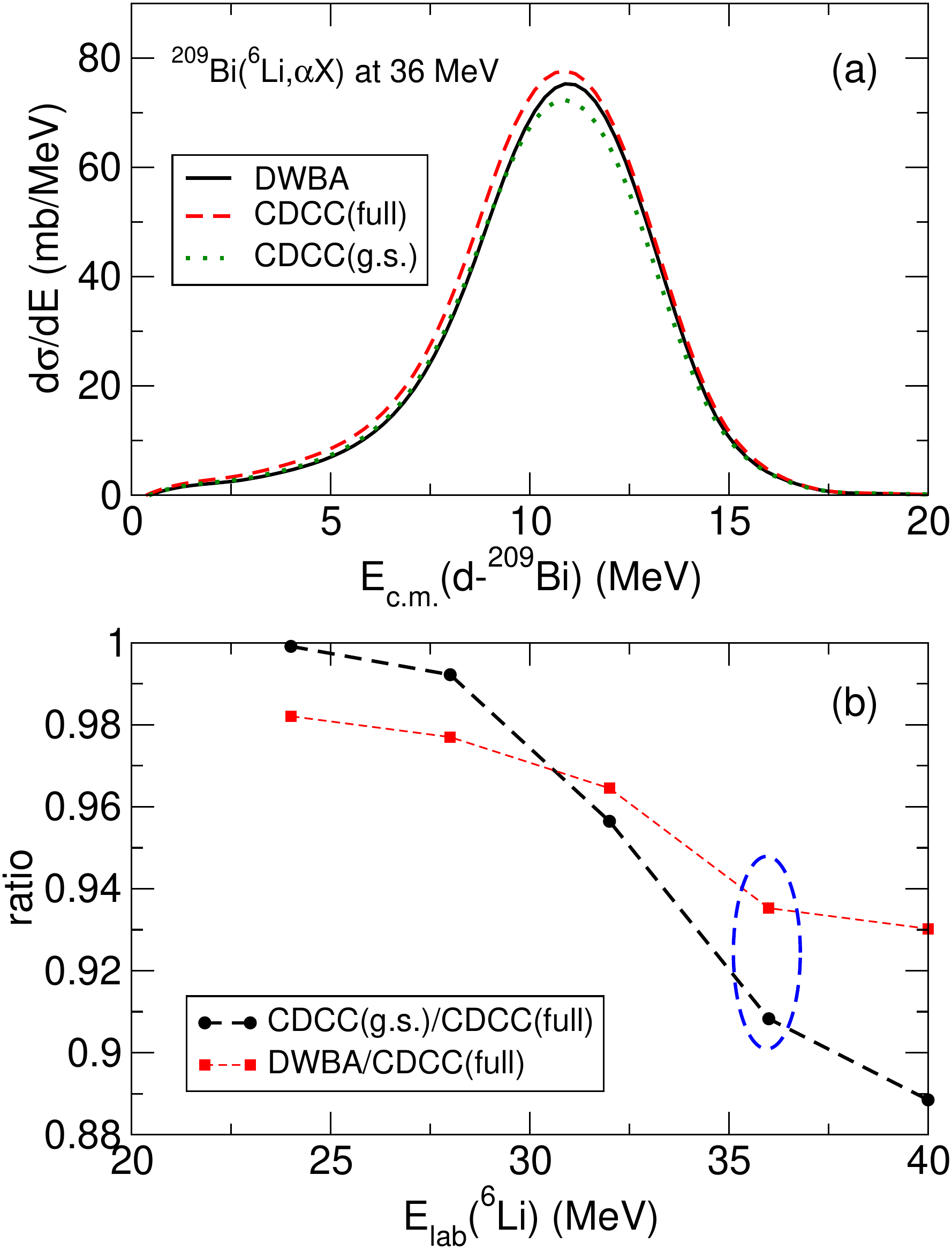}} \par}
\caption{\label{fig:li6} (a) NEB differential cross sections as a function of the $d$-$^{209}$Bi  relative  energy. (b) Ratio of NEB cross section computed by different methods as a function of the $^6$Li incident energy. The ellipse highlights the energy corresponding to  panel (a). See text for more details.}
\end{center}
\end{figure}
%-------------------------------------------------------------

As a second example, we consider the $\alpha$ production in reactions induced by the weakly bound nucleus $^6$Li.  These $\alpha$ yields are experimentally found to be very large, significantly exceeding the deuteron production channel (see e.g.~\cite{Santra12}). This result points toward NEB mechanisms, as it has been indeed confirmed by our recent calculations using the IAV model \cite{Jin15,Jin17,Jin19}. 
%The understanding of large $\alpha$ yields observed in reactions with $^6$Li has been studied in our previous work~\cite{Jin15,Jin17}. These research have shown that the total
%In our recent works \cite{Jin15,Jin17,Jin19}, we have shown that the large yields of $\alpha$ particles observed in these reactions are mostly due to  NEB modes (such as ICF).
Furthermore, the fact that a significant part of the incident flux feeds the $\alpha$-production channel results in a sizable reduction ($\sim$30\%) of the CF cross sections, as found in many experiments and confirmed by the calculations \cite{Jin19}.
% {\it In addition, the ICF is directly related to NEB. However, one has to remember that in addition to ICF contributions, other contributions, such as breakup accompanied by target excitation, are also components in NEB. Then discussion of NEB will help us understanding the path of ICF.} 

For the present study, we have considered the $^6$Li+$^{209}$Bi reaction at several energies around the Coulomb barrier ($V_b=30.1$ MeV~\cite{Das02}). Inclusive breakup data for this reaction have been compared in our previous work~\cite{Jin15} with IAV-DWBA calculations. Here, we adopt the same potentials employed in those calculations. For simplicity, we also ignore the particle spins. In the CDCC calculation, we consider the partial waves $\ell=0$-$2$ and excitation energies up to $20$ MeV for the $\alpha$-$d$ continuum. For the DWBA calculation, the $^6$Li+$^{209}$Bi potential is taken from the global parametrization of  Cook~\cite{cook82}, but we slightly adjust the potential depth to have a better agreement with the elastic scattering angular distribution obtained with CDCC.

The results are shown in Fig.~\ref{fig:li6}(a) for the angle-integrated NEB differential cross sections $\alpha$ energy distribution in the  C.M.\ frame, with the same meaning for the lines as in Fig.~\ref{fig:d93nb}. The results are qualitatively similar to those found in the deuteron case, namely, the (i) the IAV-CDCC(gs) calculation, in which  only the ground state wave function of the projectile is retained, is very close to the full calculation and  (ii) the IAV-DWBA approximation provides a good approximation to the full three-body IAV-CDCC result.  Thus, also in this reaction we find that the NEB processes proceed directly from the $^6$Li ground state. {In the case of the ICF channels, this means that the deuteron is directly captured by the target nucleus, without requiring the previous dissociation of the $^{6}$Li projectile into $\alpha+d$.}

In Fig.~\ref{fig:li6}(b) we compare the ratio of these calculations for different $^6$Li incident energies. The circles are the ratio between the IAV-CDCC(gs) and full IAV-CDCC results and the squares give the ratio between IAV-DWBA and IAV-CDCC. The dashed ellipse highlights the results of Fig.~\ref{fig:li6}(a).  It is seen that the omission of the $\alpha+d$ breakup channels (as done in the IAV-DWBA and IAV-CDCC(gs)) results in an underestimation of the NEB yield and that this effect increases with increasing incident energies. This result can be understood as due to the increasing importance of the projectile dissociation  as the incident energy increases. At the maximum incident energy explored in our calculations, the  omission of the two-step  mechanism results in a difference of 11\% in the evaluated NEB cross section. We see also in Fig.~\ref{fig:li6}(b) that the  IAV-DWBA calculation is rather close to the full IAV-CDCC calculation. As the incident energy increases, the difference with IAV-CDCC is smaller than in the case of IAV-CDCC(gs) (7\% at $E=40$~MeV), indicating the ability of the DWBA approximation of implicitly accounting for the projectile dissociation. 
%DWBA and CDCC(g.s.) slightly deviate from the full IAV-CDCC result. This can be understood as in increasing importance of the breakup channels with increasing incident energy. However, the difference is rather small. At 40~MeV, the IAV-DWBA differs by only 7\% and the IAV-CDCC(g.s.) by 11\%.. These results reinforces the reliability of DWBA. \sout{and show that the NEB is mainly coming from $^6$Li ground state. }
%-------------------------------------------------------------
\begin{figure}[tb]
\begin{center}
 {\centering \resizebox*{0.8\columnwidth}{!}{\includegraphics{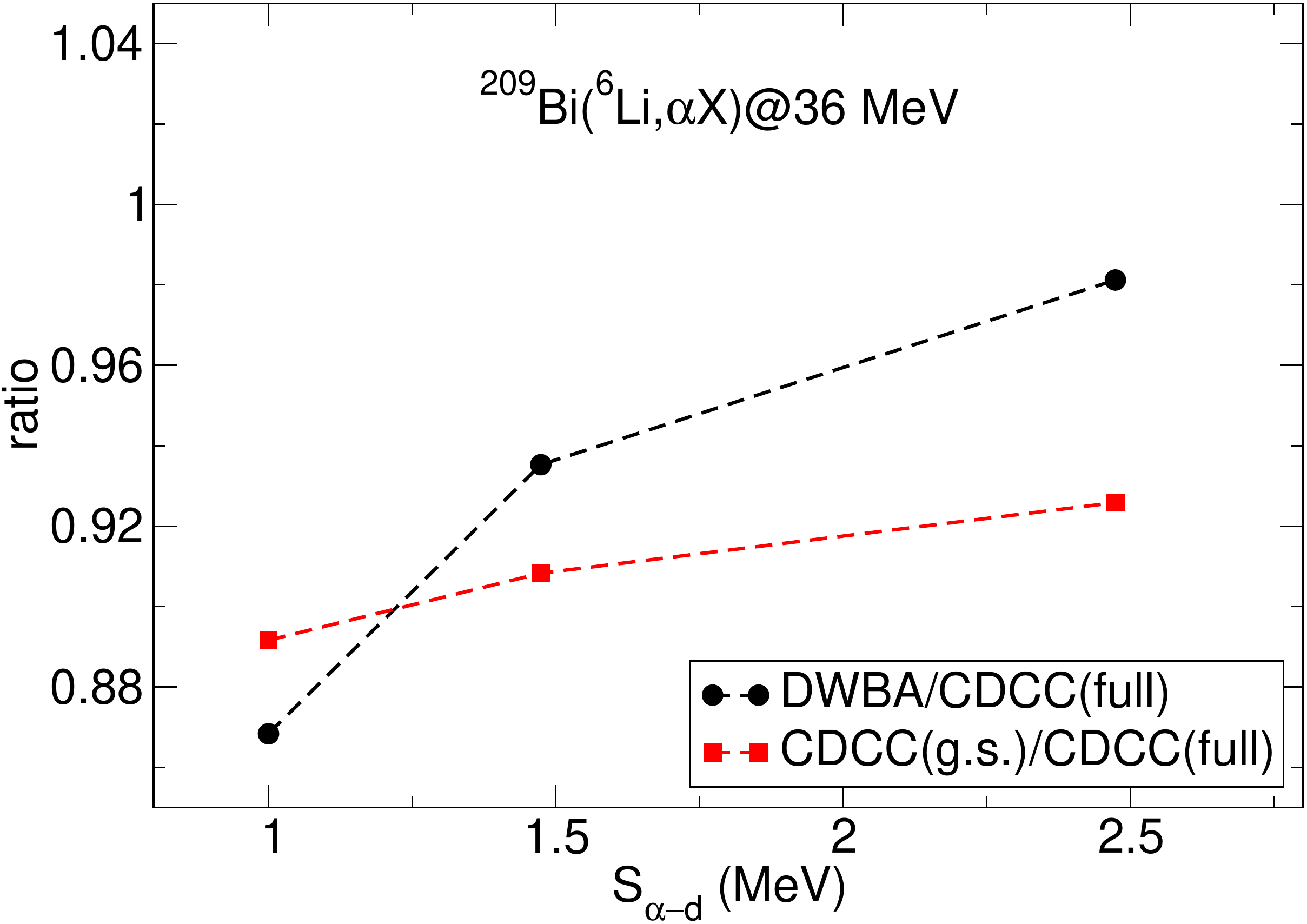}} \par}
\caption{\label{fig:ebind} Ratios of NEB cross section for the $\alpha$-production channel in the $^6$Li+$^{209}$Bi reaction as a function of the $^6\text{Li} \to \alpha+d$ separation energy.}
\end{center}
\end{figure}
%------------------------------------------------------------

%Although breakup fusion process plays a minor role in NEB, the importance of ICF and its effect on the CF suppression is known to be  closely correlated with the separation energy of the projectile \cite{Gas09,Jin19}. We investigate also here  dependence of 

The projectile dissociation (corresponding to the first step in Eq.~(\ref{eq:2step})) is known to be correlated with the separation energy of the projectile, becoming more important as the binding energy decreases. Thus, it is expected that the importance of the two-step mechanism will be also correlated with the separation energy.  To investigate this connection within the present framework, we have repeated the NEB calculations varying artificially the separation energy of $^6$Li for the  $^6$Li+$^{209}$Bi reaction at $36$ MeV. The results are shown in Fig.~\ref{fig:ebind}. The symbols have the same meaning as in Fig.~\ref{fig:li6} (b). 
%{\it It is seen that the DWBA provides an overall good agreement compare to CDCC and the ground state contribution decreases with the separation energy. } 
These results show, as expected, that IAV-CDCC(g.s.) approaches the full IAV-CDCC  when the separation energy increases. For the most weakly bound case considered in our calculations ($S_{\alpha d}=1$~MeV), the NEB cross section computed with CDCC(g.s.) underestimates by $\sim$11\%  the full IAV-CDCC result, confirming the increasing relevance of the projectile dissociation for weakly bound nuclei. The IAV-DWBA follows a similar trend compared to IAV-CDCC(gs), although the differences with IAV-CDCC are smaller except for the most weakly bound case. 

%\bigskip
%--------------------------------
{\it Summary and conclusions}.--
 \label{sec:sum}
%---------------------------------
In summary, we have presented the first implementation of the IAV model for the inclusive breakup of two-body projectiles, using a full three-body description of the scattering problem. For that, we have employed the CDCC model wavefunction. This implementation goes beyond the DWBA approximation  employed so far in previous applications of this model. 

In the range of energies explored here, however, differences remain of the order of 10\% or less, which seems to explain the success of the DWBA to account for experimental data \cite{Jin15,Jin18,Pot17,Carlson2016}. 

We have also explored the importance of the two-step process in the NEB mechanism, by comparing the full IAV-CDCC results with those obtained retaining only the projectile g.s.\ in the evaluation of the NEB cross section.  
We find that, as the separation energy decreases, or the incident energy increases,  the IAV-CDCC(g.s.) tends to deviate from the full IAV-CDCC results. Yet, the overall effect is rather small for all explored incident and binding energies (less than 12\%). 
%This is an important result because it contradicts the commonly assumed breakup-fusion picture. 
Instead, our present results conclusively show that the partial fusion process (i.e.~ICF) is mainly a one-step process and that the two-step mechanism, while not completely negligible, represents a minor contribution. These results put into question the commonly accepted breakup-fusion picture of the ICF process.

%tends to the full IAV-CDCC result wheresults are rather close to the full IAV-CDCC. latter deviates from the full IAV-CDCC results    
% In addition, when only employ the ground state part of CDCC wave function, the calculated results are closed to the full solutions. This 
\bigskip
%----------------------
\begin{acknowledgments}
%----------------------
We are grateful to Joaqu\'in G\'omez-Camacho, Gerard Baur and Mahananda Dasgupta for a critical reading of the manuscript. 
This work has been partially supported by the National Science Foundation
under Contract No.\ NSF-PHY-1520972 with Ohio University,
by the Spanish Ministerio de Ciencia, Innovaci\'on y Universidades and FEDER funds under project FIS2017-88410-P  and by the European Union's Horizon 2020 research and innovation program under Grant Agreement No.\ 654002.
 This research used resources of the National Energy Research Scientific Computing Center (NERSC), a U.S. Department of Energy Office of Science User Facility operated under Contract No. DE-AC02-05CH11231. Finally, J.L. wants to thank, in particular, the invaluable support and encouragement received from Ofelia Liu over years. 
\end{acknowledgments}

\bibliography{inclusive_prc.bib}
\end{document}